\documentclass[runningheads,amssymbols]{llncs}
\usepackage{color}
\usepackage{multirow}
\usepackage{soul}

\title{CPBPV: A Constraint-Programming Framework \\ For Bounded Program Verification}

\titlerunning{A Constraint-Programming Framework for Bounded Program Verification }

\author{H\'el\`ene Collavizza\inst{1},   Michel Rueher\inst{1}, Pascal  Van Hentenryck\inst{2}}

\institute{
  Universit\'e de Nice--Sophia Antipolis, France (\email{\{helen,rueher\}@polytech.unice.fr})\and
  Brown University, Box 1910, Providence, RI 02912 (\email{pvh@cs.brown.edu})
}

\begin{document}

\maketitle

\begin{abstract}
  This paper studies how to verify the conformity of a program with
  its specification and proposes a novel constraint-programming
  framework for bounded program verification (CPBPV). The CPBPV
  framework uses constraint stores to represent the specification and
  the program and explores execution paths nondeterministically.  The
  input program is partially correct if each constraint store so
  produced implies the post-condition.  CPBPV does not explore
  spurious execution paths as it incrementally prunes execution paths
  early by detecting that the constraint store is not consistent.
  CPBPV uses the rich language of constraint programming to express
  the constraint store. Finally, CPBPV is parametrized with a list of
  solvers which are tried in sequence, starting with the least
  expensive and less general. Experimental results often produce
  orders of magnitude improvements over earlier approaches, running
  times being often independent of the variable domains. Moreover,
  CPBPV was able to detect subtle errors in some programs while other
  frameworks based on model checking have failed.
\end{abstract}

\section{Introduction}

This paper is concerned with software correctness, a critical issue in
software engineering. It proposes a novel constraint-programming
framework for bounded program verification (CPBPV), i.e., when the
program inputs (e.g., the array lengths and the variable values) are
bounded.  The goal is to verify the conformity of a program with its
specification, that is to demonstrate that the specification is a
consequence of the program.  The key idea of CPBPV is to use
constraint stores to represent the specification and the program, and
to non-deterministically explore execution paths over these constraint
stores. This non-deterministic constraint-based symbolic execution
incrementally refines the constraint store, which initially consists
of the precondition. Non-determinism occurs when executing conditional 
or iterative instructions and the non-deterministic execution refines
the constraint store by adding constraints coming from conditions and
from assignments. The input program is partially correct if each
constraint store produced by the symbolic execution implies the
post-condition. It is important to emphasize that CPBPV considers
programs with complete specifications and that verifying the
conformity between a program and its specification requires to check
(explicitly or implicitly) all executables paths. This is not the case
in model-checking tools designed to detect violations of some specific
property, e.g., safety or liveness properties.

The CPBPV framework has a number of fundamental benefits.  First, contrary
to earlier work using constraint programming or SMT
\cite{ABM07,CoR06,CoR07}, CPBPV does not use predicate abstraction or
explore spurious execution paths, i.e., paths that do not correspond
to actual executions over inputs satisfying the pre-condition.  CPBPV
incrementally prunes execution paths early by detecting that the
constraint store is not consistent. Second, CPBPV uses the rich
language of constraint programming to express the constraint store,
including arbitrary logical and threshold combination of constraints,
the {\em element} constraint, and global/combinatorial constraints
that express complex relationships on a set of variables. Finally,
CPBPV is parametrized with a list of solvers which are tried in
sequence, starting with the least expensive and less general.

The CPBPV framework was evaluated experimentally on a series of
benchmarks from program verification.  Experimental results of our
(slow) prototype often produce orders of magnitude improvements over
earlier approaches, and indicate that the running times are often
independent of the variable domains. Moreover, CPBPV was able to found
subtle errors in some programs that some other verification frameworks
based on model-checking could not detect.

The rest of the paper is organized as follows. Section
\ref{motivation} illustrates how CPBPV handles constraints store on a
motivating example.  Section \ref{formalization} formalizes the CPBPV
framework for a small programming language and Section
\ref{implementation} discusses the implementation issues. Section
\ref{experimental} presents experimental results on a number of
verification problems, comparing our approach with state of the art
model-checking based verification frameworks.  Section \ref{related}
discusses related work in test generation, bounded program
verification and software model checking. Section \ref{conclusion}
summarizes the contributions and presents future research directions.

\section{The Constraint-Programming Framework at Work} 
\label{motivation}

This section illustrates the CPBPV verifier on a motivating example,
the binary search program.  CPBPV uses Java programs and JML
specifications for the pre- and post-conditions, appropriately
enhanced to support the expressivity of constraint programming. Figure
\ref{BsearchFig} depicts a binary search program to determine if a
value $v$ is present in a sorted array $t$. (Note that
$\backslash${\tt result} in JML corresponds to the value returned by
the program).  To verify this program, our prototype implementation
requires a bound on the length of array $t$, on its elements, and on
$v$. We will verify its correctness for specific lengths and simply
assume that the values are signed integers on a number of bits.

\begin{figure}[t]
{\footnotesize
\begin{verbatim}
/*@ requires (\forall int i; i>=0 && i<t.length-1;t[i]<=t[i+1])
  @ ensures
  @   (\result != -1 ==> t[\result] == v) &&
  @   (\result == -1 ==> \forall int k; 0 <= k < t.length ; t[k] != v) @*/
1 static int binary_search(int[] t, int v) {
2    int l = 0;
3    int u = t.length-1;
4    while (l <= u) {
5      int m = (l + u) / 2;
6      if (t[m]==v) 
7          return m;
8      if (t[m] > v) 
9         u = m - 1;
10      else 
11        l = m + 1;    }   // ERROR else u = m - 1;
12   return -1; }
\end{verbatim}}
\vspace{-0.4cm}
\caption{The Binary Search Program}
\label{BsearchFig}
\vspace{-0.4cm}
\end{figure}

The initial constraint store of the CPBPV verifier, assuming an input
array of length 8, is the precondition\footnote{We omit the domain
  constraints on the variables for simplicity.}
$c_{pre} \equiv \forall 0 \leq i < 7: t^0[i] \leq t^0[i+1]$ 
where $t^0$ is an array of constraint variables capturing the
input. The constraint variables are annotated with a version number as
CPBPV performs a SSA-like renaming \cite{CFR91} on the fly since each
assignment generates constraints possibly linking
the old and the new values of the assigned variable. The assignments
in lines 2--3 add the constraints $l^0 = 0 \wedge u^0 = 7$. CPBPV then
considers the loop instruction. Since $l^0 \leq u^0$, it enters the
loop body, adds the constraint $m^0 = (l^0 + u^0)/2$, which simplifies
to $m^0 = 3$, and considers the conditional statement on line 6.  The
execution of the statement is nondeterministic: Indeed, both $t^0[3]
 = v^0$ and $t^0[3] \neq v^0$ are consistent with the constraint store,
so that the two alternatives, which give rise to two execution paths,
must be explored. Note that these two alternatives correspond to
actual execution paths in which $t[3]$ in the input is equal to, or
different from, input $v$. The first alternative adds the constraint
$t^0[3] = v^0$ to the store and executes line 7 which adds the
constraint $result = m^0$. CPBPV has thus obtained an execution path
$p$ whose final constraint store $c_p$ is:
\begin{small}
$
c_{pre}
\; \wedge \; l^0 = 0 \wedge u^0 = 7
\; \wedge \; m^0 = (l^0 + u^0)/2
\; \wedge \; t^0[m^0] = v^0
\; \wedge \; result = m^0
$\\
\end{small}
CPBPV  then checks whether this store $c_p$ implies the
post-condition $c_{post}$ by searching for a solution to $c_p \;
\wedge \; \neg c_{post}$. This test fails, indicating that the
computation path $p$, which captures the set of actual executions in
which $t[3] = v$, satisfies the specification. CPBPV then
explores the other alternatives to the conditional statement in line
6. It adds the constraint $t^0[m^0] \neq v^0$ and executes the
conditional statement in line 8. Once again, this statement is
nondeterministic. Its first alternative assumes that the test holds,
generating the constraint $t^0[m^0] > v^0$ and executing the
instruction in line 9. Since $u$ is (re-)assigned, CPBPV 
creates a new variable $u^1$ and posts the constraint $u^1 = m^0 -
1 =2$. 
The execution returns to line 4, where the test now reads $l^0
\leq u^1$, since CPBPV  always uses the most recent
version for each variable. Since the constraint stores entails $l^0
\leq u^1$, the only extension to the current path consists of
executing line 5, adding the constraint $m^1 = (l^0 + u^1)/2$, which
actually simplifies to $m^1 = 1$. Another complete execution path is
then obtained by executing lines 6 and 7. 


Consider now a version of the program in which line 11 is replaced by
{\tt u = m-1}. To illustrate the CPBPV verifier, we specify partial
execution paths by indicating which alternative is selected for each
nondeterministic instruction. For instance, $\langle
T_4,F_6,T_8,T_5,T_6\rangle$ denotes the last execution path discussed
above in which the true alternative is selected for the first
execution of the instruction in line 4, the false alternative for the
first execution of instruction 6, the true alternative for the first
instruction of instruction 8, the true alternative of the second
execution of instruction 5, and the true alternative of the second
execution of instruction 6. Consider the partial path $\langle
T_4,F_6,F_8 \rangle$ and let us study how it can be extended. 
The partial path $\langle T_4,F_6,F_8,T_4,T_6 \rangle$ is not explored,
since it produces a constraint store containing \\
\hspace*{3cm}$
c_{pre}
\; \wedge \; t^0[3] \neq v^0
\; \wedge \; t^0[3] \leq v^0
\; \wedge \; t^0[1] = v^0
$\\
which is clearly inconsistent. Similarly, the path $\langle
T_4,F_6,F_8,T_4,F_6,T_8\rangle$ cannot be extended. The output of
CPBPV on this incorrect program when executed on an array of length 8
(with integers coded on 8-bits to make it readable) produces, in 0.025
seconds, the counterexample:\\
\begin{small}
$
v^0 = -126 \ \wedge \ t^0 = [-128,-127,-126,-125,-124,-123,-122,-121] \ \wedge \ result = -1.
$\\
\end{small}

\noindent
This example highlights a few interesting benefits of CPBPV.
\begin{enumerate}
\item The verifier only considers paths that correspond to collections
  of actual inputs (abstracted by constraint stores). The resulting
  execution paths must all be explored since our goal is to prove the
  partial correctness of the program.
  
\item The performance of the verifier is independent of the integer
  representation on this application: it only requires a bound on the
  length of the array.

\item The verifier returns a counter-example for debugging the
  program.
\end{enumerate}
Note that $CBMC$ and $ESC/Java 2$, two state-of-the-art
model checkers 
fail to verify this example as discussed in Section
\ref{experimental}.

\section{Formalization of the Framework} 
\label{formalization}
\label{semantics}

This section formalizes the CPBPV verifier on a small abstract
language using a small-step SOS semantics. The semantics primarily
specifies the execution paths over constraint stores explored by the
verifier. It features \verb+assert+ and \verb+enforce+ constructs
which are necessary for modular composition.

\paragraph{\bf  Syntax}

Figure \ref{syntax-c} depicts the syntax of the programs and the
constraints generated by the verifier. In the following, we use $s$,
possibly subscripted, to denote elements of a syntactic entity $S$.

\newcommand{\brangle}{\big\rangle}
\newcommand{\blangle}{\big\langle}
{\small
\begin{figure}[t]
\begin{footnotesize}
\begin{eqnarray*}
\begin{array}{l}
L: \mbox{\it list of instructions}; I: {\it instructions}; B: \mbox{\it Boolean expressions} \\
E: \mbox{\it integer expressions}; A: {\it arrays}; V: \mbox{\it variables} \\
\\
L ::= I ; L \; | \; \epsilon \\
I ::= A[E] \leftarrow E \; | \; V \leftarrow E \; | \; {\bf\it if} \ B \ I  \; | \; {\bf\it while} \ B \ I \; | \; {\bf\it assert(B)} \; | \; {\bf\it enforce(B)} \; | \;
      {\bf\it return} \ E \; | \; \{ L \} \\
B ::= true \;|\; false \;|\; E > E \;| \; E \geq E \;| \; E = E \;| \; E \neq E \;| \; E \leq E \;| \; E < E  \\
B ::= \neg B \;| \; B \wedge B \; | \; B \vee B \;| \; B \Rightarrow B \\
E ::= V \;|\; A[E] \;|\; E + E \;|\; E - E \;|\; E \times E \;|\; E / E \;|\;
\\ \\ 
C: \mbox{\it constraints} \hspace*{2cm}
E^+: \mbox{\it solver expressions} \\
V^+ = \{ v^i \ | \ v \in V \ \& \ i \in {\cal N}\}:  \mbox{\it solver variables} \\
A^+ = \{ a^i \ | \ a \in A \ \& \ i \in {\cal N}\}:  \mbox{\it solver arrays} \\
\\
C ::= true \;|\; false \;|\; E^+ > E^+ \;| \; E^+ \geq E^+ \;| \; E^+ = E^+ \;| \; E^+ \neq E^+ \;| \; E^+ \leq E^+ \;| \; E^+ < E^+ \\
C ::= \neg C \;| \; C \wedge C \; | \; C \vee C \;| \; C \Rightarrow C \\
E^+ ::= V \;|\; A[E^+] \;|\; E^+ + E^+ \;|\; E^+ - E^+ \;|\; E^+ \times E^+ \;|\; E^+ / E^+ \;|\;
\end{array}
\end{eqnarray*}
\end{footnotesize}
\vspace{-0.5cm}
\caption{The Syntax of Programs and Constraints}
\label{syntax-c}
\vspace{-0.5cm}
\end{figure}
}
\paragraph{\bf Renamings} 

CPBPV creates variables and arrays of variables ``on-the-fly'' when
they are needed. This process resembles an SSA normalization but does
not introduce the join nodes, since the results of different execution
paths are not merged. Similar renamings are used in model
checking. The renaming uses mappings of type $V \cup A \rightarrow
{\cal N}$ which maps variables and arrays into a natural numbers
denoting their current ``version numbers''. In the semantics, the
version number is incremented each time a variable or an array element
is assigned. We use $\sigma_{\bot}$ to denote the uniform mapping to
zero (i.e., $\forall x \in V \cup A: \sigma_{\bot}(x) = 0$) and
$\sigma[x/i]$ the mapping $\sigma$ where $x$ now maps to $i$, i.e.,
$\sigma[x/i](y) = {\it if} x = y \mbox{ {\it then} } i \mbox{ {\it else} } \sigma(y).$
These mappings are used by a polymorphic renaming function $\rho$ to
transform program expressions into constraints.  For example, $\rho \
\sigma \ b_1 \oplus b_2 = (\rho \ \sigma \ b_1) \oplus (\rho \ \sigma
\ b_2) (\mbox{where } \oplus \in \{\wedge,\vee,\Rightarrow\})$ is the
rule used to transform a logical expression.

\paragraph{\bf Configurations}

The CPBCV semantics mostly uses configurations of the type $\langle l,
\sigma, c \rangle$, where $l$ is the list of instructions to execute,
$\sigma$ is a version mapping, and $c$ is the set of constraints
generated so far. It also uses configurations of the form $\langle
\top, \sigma, c \rangle$ to denote final states and configurations of the form
$\langle \bot, \sigma, c \rangle$ to denote the violation of an
assertion. The semantics is specified by rules of the form
$
\frac{\mbox{conditions}}
{\gamma_1  \longmapsto \gamma_2}
$
stating that configuration $\gamma_1$ can be rewritten into
$\gamma_2$ when the conditions hold.

\paragraph{\bf Conditional Instructions} The conditional instruction
${\bf\it if} \ b \ i$ considers two cases. If the constraint $c_b$
associated with $b$ is consistent with the constraint store, then the
store is augmented with $c_b$ and the body is executed. If the
negation $\neg c_b$ is consistent with the store, then the constraint
store is augmented with $\neg c_b$. Both rules may apply, since the
store may represent some memory states satisfying the condition and
some violating it.

\begin{minipage}[t]{1.8in}
\[
\frac{c \wedge (\rho \ \sigma \ b) \mbox{ is satisfiable}}
{\langle {\bf\it if} \ b \ i \; ; \; l, \sigma, c \rangle \longmapsto \langle i \; ; \; l, \sigma,  c \wedge (\rho \ \sigma \ b) \rangle}
\]
\end{minipage}
$\;\;\; \;\;\;$ \hfill
\begin{minipage}[t]{2.2in}
\[
\frac{c \wedge \neg (\rho \ \sigma \ b) \mbox{ is satisfiable}}
{\langle {\bf\it if} \ b \ i \; ; \; l, \sigma, c \rangle \longmapsto \langle l, \sigma,  c \wedge \neg (\rho \ \sigma \ b) \rangle}
\]
\end{minipage}

\paragraph{\bf Iterative Instructions} The while instruction ${\bf\it
  while} \ b \ i$ also considers two cases. If the constraint $c_b$
associated with $b$ is consistent with the constraint store, then the
constraint store is augmented with $c_b$, the body is executed, and
the while instruction is reconsidered. If the negation $\neg c_b$ is
consistent with the constraint store, then the constraint store is
augmented with $\neg c_b$. 


\[
\frac{c \wedge (\rho \ \sigma \ b) \mbox{ is satisfiable}}
{\langle {\bf\it while} \ b \ i \; ; \; l, \sigma, c \rangle \longmapsto \langle i;while \ b \ i \; ; \; l, \sigma,  c \wedge (\rho \ \sigma \ b) \rangle}
\]
\[
\frac{c \wedge \neg (\rho \ \sigma \ b) \mbox{ is satisfiable}}
{\langle {\bf\it while} \ b \ i \; ; \; l, \sigma, c \rangle \longmapsto \langle l, \sigma,  c \wedge \neg (\rho \ \sigma \ b) \rangle}
\]

\paragraph{\bf Scalar Assignments} Scalar assignments create a new
constraint variable for the program variable to be assigned and add a
constraint specifying that the variable is equal to the right-hand
side.  A new renaming mapping is produced.

\[
\frac{\sigma_2 = \sigma_1[v/\sigma_1(v)+1] \;\; \& \;\; 
c_2 \equiv (\rho \ \sigma_2 \ v) = (\rho \ \sigma_1 \ e)}
{\langle v \leftarrow e \; ; \; l, \sigma_1, c_1 \rangle \longmapsto \langle l, \sigma_2, c_1 \wedge c_2 \rangle}
\]

\paragraph{\bf Assignments of Array Elements} The assignment of an
array element creates a new constraint array, add a constraint for the
index being indexed and posts constraints specifying that all the new
constraint variables in the array are equal to their earlier version,
except for the element being indexed. Note that the index is an
expression which may contain variables as well, giving rise to the
well-known {\it element} constraint in constraint programming
\cite{VanHentenryck89}.

\[
\frac{\begin{array}{l}
\sigma_2 = \sigma_1[a/\sigma_1(a)+1] \\
c_2 \equiv (\rho \ \sigma_2 \ a) [\rho \ \sigma_1 \ e_1]  = (\rho \ \sigma_1 \ e_2) \\
c_3 \equiv \forall i \in 0..{\it a.length}:  (\rho \ \sigma_1 \ e_1) \neq i \; \Rightarrow \; (\rho \ \sigma_2 \ a) [i] = (\rho \ \sigma_1 \ a) [i] 
\end{array}}
{\langle a[e_1] \leftarrow e_2, \sigma_1\; ; \; l, c_1 \rangle \longmapsto \langle l, \sigma_2, c_1 \wedge c_2 \wedge c_3 \rangle}
\]

\paragraph{\bf Assert Statements} An assert statement checks whether the assertion
is implied by the control store in which case it proceeds normally. Otherwise,
it terminates the execution with an error.

\begin{minipage}[t]{1.8in}
\[
\frac{
c  \Rightarrow (\rho \ \sigma \ b)
}
{\langle {\bf\it assert} \; b \; ; \; l, \sigma, c \rangle \longmapsto \langle 
l, \sigma, c \rangle}
\]
\end{minipage}
$\;\;\;$ \hfill
\begin{minipage}[t]{1.8in}
\[
\frac{
c \wedge \neg (\rho \ \sigma \ b) \mbox{ is satisfiable}
}
{\langle {\bf\it assert} \; b \; ; \; l, \sigma, c \rangle \longmapsto \langle 
\bot, \sigma, c \rangle}
\]
\end{minipage}

\paragraph{\bf Enforce Statements} An enforce statement adds a constraint to the constraint
store if it is satisfiable.
\[
\frac{
c \wedge (\rho \ \sigma \ b) \mbox{ is satisfiable}
}
{\langle {\bf\it enforce} \; b \; ; \; l, \sigma, c \rangle \longmapsto \langle 
l, \sigma, c \wedge (\rho \ \sigma \ b) \rangle}
\]

\paragraph{\bf Block Statements} Block statements simply remove the braces.
\[
{\langle \{ l_1 \}\; ; \; l_2, \sigma, c \rangle \longmapsto \langle l_1:l_2, \sigma, c \rangle}
\]

\paragraph{\bf Return Statements} A return statement simply constrains the {\it result} variable.
\[
\frac{
c_2 \equiv (\rho \ \sigma_1 \ result) = (\rho \ \sigma_1 \ e) 
}
{\langle {\bf\it return} \ e \; ; \; l, \sigma_1, c_1 \rangle \longmapsto \langle \sigma_1, c_1 \wedge c_2 \rangle}
\]

\paragraph{\bf Termination} Termination also occurs when no instruction remains.
\[
{\langle \epsilon, \sigma, c \rangle \longmapsto \langle \top, \sigma, c \rangle}
\]

\paragraph{\bf The CPBPV Semantics} Let ${\cal P}$ be program 
$b_{pre} \; l \; b_{post}$ in which $b_{pre}$ denotes the
precondition, $l$ is a list of instructions, and $b_{post}$ the
post-condition. Let $\stackrel{*}{\longmapsto}$ be the transitive
closure of $\longmapsto$. The final states are specified by the set
\[
{\it SFN}(b_{pre},{\cal P}) = \{ \ \langle f, \sigma, c \rangle | 
\langle i, \sigma_{\bot}, \rho \ \sigma_{\bot} \ b_{pre} \rangle \stackrel{*}{\longmapsto}{*} \langle f , \sigma, c \rangle \; \wedge \; f \in \{\bot,\top\} \ \}
\]
The program violates an assertion if the set 
\[
{\it SFE}(b_{pre},{\cal P},b_{post}) = \{ \langle \bot, \sigma, c \rangle \in {\it SFN}(b_{pre},{\cal P}) \}
\]
is not empty. It violates its specification if the set
\[ 
{\it SFE}(b_{pre},{\cal P},b_{post}) = \{ \top, \sigma, c \rangle \in {\it SFN}(b_{pre},{\cal P}) \ | \ c \ \wedge \ (\rho \ \sigma \ \neg b_{post}) \mbox{ satisfiable} \}
\]
is not empty. It is partially correct otherwise.

\section{Implementation issues} 
\label{implementation}

The CPBPV framework is parametrized by a list of solvers
$(S_1,\ldots,S_k)$ which are tried in sequence, starting with the
least expensive and less general. When checking satisfiability, the
verifier never tries solver $S_{i+1},\ldots,S_{k}$ if solver $S_i$ is
a decision procedure for the constraint store. If solver $S_i$ is not
a decision procedure, it uses an abstraction $\alpha$ of the constraint
store $c$ satisfying $c \Rightarrow \alpha$ and can still detect
failed execution paths quickly. The last solver in the sequence is a
constraint-programming solver (CP solver) over finite domains which iterates
pruning and searching to find solutions or prove infeasibility.  When
the CP solver makes a choice, the earlier solvers in the sequence are
called once again to prune the search space or find solutions if they
have become decision procedures. Our prototype implementation uses a
sequence $(MIP,CP)$, where MIP is the mixed integer-programming tool
ILOG CPLEX\footnote{See http://www.ilog.com/products.}  and CP is the
constraint-programming tool Ilog JSOLVER.  Our Java implementation
also performs some trivial simplifications such as constant propagation
but is otherwise not optimized in its use of the solvers and in its
renaming process whose speed and memory usage could be improved
substantially. 
Practically, simplifications are done on  the fly and  the
  MIP solver is  called at each node of the executable paths. The CP
  solver is only called at the end  of the executable paths when the
  complete post condition is considered.
Currently, the implementation use a depth-first strategy 
  for the CP  solver, but modern CP languages now offer high-level abstractions
to implement other exploration strategies. In practice, when CPBPV
is used for model checking as discussed below, it is probably
advisable to use a depth-first iterative deepening implementation.

\section{Experimental results} 
\label{experimental}

In this section, we report experimental results for a set of
traditional benchmarks for program verification.  We compare CPBVP
with the following frameworks:

\begin{itemize}

\item ESC/Java is an Extended Static Checker for Java to find common
  run-time errors in JML-annotated Java programs by static analysis of
  the code and its annotations. See
  http://kind.ucd.ie/products/opensource/ESCJava2/.

\item CBMC is a Bounded Model Checker for ANSI-C and C++ programs. It
  allows for the verification of array bounds (buffer overflows),
  pointer safety, exceptions, and user-specified assertions. See
  http://www.cprover.org/cbmc/.

\item BLAST, the Berkeley Lazy Abstraction Software Verification Tool,
  is a software model checker for C programs. See
  http://mtc.epfl.ch/software-tools/blast/.

\item EUREKA is a C bounded model checker which uses an SMT solver
  instead of an SAT solver. See http://www.ai-lab.it/eureka/.

\item Why is a software verification platform which integrates many
  existing provers (proof assistants such as Coq, PVS, HOL 4,...) and
  decision procedures such as Simplify, Yices, ...). See
  http://why.lri.fr/.
\end{itemize}

\noindent
Of course, neither the expressiveness nor the objectives of
all these systems are the same as the one of CPBPV.  For instance,
some of them can handle CTL/LTL constraints whereas CPBPV dos not yet
support this kind of constraints. Nevertheless, this comparison is
useful to illustrate the capabilities of CPBPV.   

\noindent
All experiments were performed on the same machine, an Intel(R)
Pentium(R) M processor 1.86GHz with 1.5G of memory, using the version
of the verifiers that can be downloaded from their web sites (except
for EUREKA for which the execution times given in \cite{ABM07,AMP06}
are reported.) For each benchmark program, we describe the data
entries and the verification parameters. In the tables, ``UNABLE''
means that the corresponding framework is unable to validate the
program either because a lack of expressiveness or because of time or
memory limitations, ``NOT\_FOUND'' that it does not detect an error,
and ``FALSE\_ERROR'' that it reports an error in a correct program.
Complete details of the experiments, including input files and error
traces, can be found in \cite{CRV08}.

\paragraph{\bf Binary search}

We start with the binary search program presented in figure
\ref{BsearchFig}.  ESC/Java is applied on the program described in
Figure \ref{BsearchFig}. ESC/Java requires a limit on the number of
loop unfoldings, which we set to $log(n)+1$ which is the worst case
complexity of binary search algorithm for an array of length $n$.
Similarly, CBMC requires an overestimate of the number of loop
unfoldings. Since CBMC does not support first-order expressions such
as JML $\setminus forall$ statement, we generated a C program for each
instance of the problem (i.e., each array length).  For example, the
postcondition for an array of length $8$ is given by

{\scriptsize
\begin{verbatim}
 (result!=-1 && a[result]==x)||
 (result==-1 && (a[0]!=x&&a[1]!=x&&a[2]!=x&&a[3]!=x&&a[4]!=x&&a[5]!=x&&a[6]!=x&&a[7]!=x)
\end{verbatim}}

\noindent
For the Why framework, we used the binary search version given in
their distribution. This program uses an assert statement to give a
loop invariant.

Note that CPBPV does not require any additional information: no
invariant and no limits on loop unfoldings. During execution, it
selects a path by nondeterministically applying the semantic rules for
conditional and loop expressions. 


Table \ref{tabsearch} reports the experimental results.  Execution
times for CPBPV are reported as a function of the array length for
integers coded on 31 bits.\footnote{The commercial MIP solver fails
  with 32-bit domains because of scaling issues.} Our implementation
is neither optimized for time or space at this stage and times are
only given to demonstrate the feasibility of the CPBPV verifier.

The ``Why'' framework \cite{FiM07} was unable to verify the
correctness without the loop invariant; 60\% of the proof obligations
remained unknown.

The CBMC framework was not able to do the verification for an instance 
of length 32 (it was interrupted after 6691,87s).

ESC/Java was unable to verify the correctness of this
program unless complete loop invariants are provided
\footnote{a version with loop invariants that allows to show the correctness
of this program has been written by David Cok, a developper of ESC/Java,
after we contacted him.}.


\begin{table}[t]
\begin{small}
\begin{center}
\begin{tabular}{{|c||l|c|c|c|c|c|c|}}
 \hline
\multirow{2}*{CPBPV} & array length & 8 & 16 & 32 & 64 & 128 & 256 \\
\cline{2-8}
        & time & 1.081s & 1.69s & 4.043s & 17.009s & 136.80s& 1731.696s \\
 \hline
\multirow{2}*{CBMC} & array length & 8 & 16 & 32 & 64 & 128 & 256 \\
\cline{2-8}
        & time & 1.37s & 1.43s & UNABLE  & UNABLE &UNABLE &UNABLE \\
 \hline
\multirow{2}*{Why} & with invariant  & \multicolumn{6}{|l|}{11.18s} \\
\cline{2-8}
                   & without invariant & \multicolumn{6}{|l|}{UNABLE} \\
 \hline
ESC/Java & \multicolumn{7}{|l|}{FALSE\_ERROR} \\
 \hline
BLAST & \multicolumn{7}{|l|}{UNABLE}\\
 \hline
\end{tabular}
\end{center}
\caption {Comparison table for binary search}\label{tabsearch}
\vspace{-0.5cm}
\end{small}
\end{table}


\paragraph{\bf An Incorrect Binary search}

Table \ref{tabsearchKO} reports experimental results for an incorrect
{\em binary search} program (see Figure \ref{BsearchFig}, line 11)
for CPBPV, ESC/Java, CBMC, and Why using an invariant. The error trace
found with CPBPV has been described in Section \ref{motivation}.  The
error traces provided by CBMC and ESC/Java only show the decisions
taken along the faulty path can be found in \cite{CRV08}. In contrast
to CPBPV, they do not provide any value for 
the array nor the searched data. Observe that CPBPV provides orders of
magnitude improvements in efficiency over CBMC and also outperforms
ESC/Java by almost a factor 8 on the largest instance.

\begin{table}[t]
\begin{small}
\begin{center}
\begin{tabular}{|c|c|c|c|c|c|c|}
 \hline
    & CPBPV & ESC/Java & CBMC   & WHY with invariant & BLAST\\
\hline
  length  8 & 0.027s & 1.21 s  & 1.38s  & NOT\_FOUND & UNABLE \\
\hline
 length  16 & 0.037s & 1.347 s  & 1.69s  & NOT\_FOUND & UNABLE \\
\hline
 length  32 &  0.064s  &  1.792 s &  7.62s  & NOT\_FOUND& UNABLE \\
\hline
 length  64 &  0.115s & 1.886 s  &  27.05s  & NOT\_FOUND& UNABLE\\
\hline
 length  128 & 0.241s & 1.964 s  &  189.20s  & NOT\_FOUND& UNABLE\\
 \hline
\end{tabular}
\end{center}
\caption{Experimental Results for an Incorrect Binary Search}
\label{tabsearchKO}
\end{small}
\end{table}

\paragraph{\bf The Tritype Program}

The tritype program is a standard benchmark in test case generation
and program verification since it contains numerous non-feasible
paths: only 10 paths correspond to actual inputs because of complex
conditional statements in the program. The program takes three
positive integers as inputs (the triangle sides) and returns 2 if the
inputs correspond to an isosceles triangle, 3 if they correspond to an
equilateral triangle, 1 if they correspond to some other triangle, and
4 otherwise. The {\tt
  tritype} program in Java with its specification in JML can be found
in\cite{CRV08}. Table 
\ref{tabTritype} depicts the experimental results for CPBPV, ESC/Java,
CBMC, BLAST and Why. BLAST was unable to validate this example because
the current version does not handle linear arithmetic. Observe the
excellent performance of CPBPV and note that our previous approach
using constraint programming and Boolean abstraction to abstract the
conditions, validated this benchmark in $8.52$ seconds when integers
were coded on 16 bits \cite{CoR07}. It also explored 92 spurious
paths.

\begin{table}[t]
\begin{small}
\begin{center}
\begin{tabular}{|c|c|c|c|c|c|}
 \hline
    & CPBPV & ESC/Java & CBMC   & Why & BLAST\\
\hline
time & 0.287s & 1.828s &  0.82s  &  8.85s & UNABLE  \\
 \hline
\end{tabular}
\end{center}
\caption{Experimental Results on the Tritype Program}
\label{tabTritype}
\end{small}
\end{table}

\paragraph{\bf An Incorrect Tritype Program}

Consider now an incorrect version of {\em Tritype} program in which
the test {\em ``if ((trityp==2)\&\&(i+k$>$j))''} in line 22 (see
\cite{CRV08})   is replaced by {\em ``if 
  ((trityp==1)\&\&(i+k$>$j))''}.  Since the local variable {\em
  trityp} is equal to {\em 2} when {\em i==k}, the condition {\em
  (i+k)$>$j} implies that {\em (i,j,k)} are the sides of an isosceles
triangle (the two other triangular inequalities are trivial because
j$>$0).  But, when {\em trityp=1}, {\em i==j} holds and this incorrect
version may answer that the triangle is isosceles while it may not be a
triangle at all.  For example, it will return {\em 2} when {\em
  (i,j,k)=(1,1,2)}. Table \ref{tabTritypeKO} depicts the experimental
results. Execution times correspond to the time required to find the
first error.  The error found with CPBPV corresponds to input values
$(i,j,k)=(1,1,2)$ mentioned earlier. Once again, observe the excellent
behavior of CPBPV compared to the remaining tools.
\footnote{For CBMC, 
we have contacted D. Kroening who has 
recommended to use the option CPROVER\_assert. If we do so, CBMC is able to find 
the error, but  we must add
some assumptions  to mean that there is no overflow into the sums,
in order to prove the correct version of tritype with this same option.}

\begin{table}[t]
\begin{small}
\begin{center}
\begin{tabular}{|c|c|c|c|c|}
 \hline
    & CPBPV & ESC/Java & CBMC    & WHY \\
\hline
time &  0.056s s & 1.853s & NOT\_FOUND  & NOT\_FOUND \\
 \hline
\end{tabular}
\end{center}
\caption{Experimental Results for the Incorrect Tritype Program}
\label{tabTritypeKO}
\end{small}
\end{table}

\paragraph{\bf Bubble Sort with initial condition}

This benchmark (see  \cite{CRV08}) is
taken from \cite{ABM07} and performs a bubble sort of an array $t$
which contains integers from $0$ to $t.length$ given in decreasing
order. Table \ref{tabbuble} shows the comparative results for this
benchmark. CPBPV was limited on this benchmark because its recursive
implementation uses up all the JAVA stack space. This problem should
be remedied by removing recursion in CPBPV.

\begin{table}[t]
\begin{small}
\begin{center}
\begin{tabular}{|c|c|c|c|c|}
 \hline
    & CPBPV & ESC/Java & CBMC  & EUREKA\\
\hline
  length  8 & 1.45s & 3.778 s & 1.11s  &  91s\\
\hline
 length  16 & 2.97s & UNABLE &  2.01s  & UNABLE\\
\hline
 length  32 &  UNABLE  & UNABLE &  6.10s  & UNABLE\\
 \hline
 length  64 &  UNABLE  & UNABLE & 37.65s  & UNABLE\\
 \hline
\end{tabular}
\end{center}
\caption{Experimental Results for Bubble Sort}
\label{tabbuble}
\end{small}
\end{table}

\paragraph{\bf Selection Sort}

We now present a benchmark to highlight both modular verification and
the {\tt element} constraint of constraint programming to index arrays
with arbitrary expressions. The benchmark  described in  \cite{CRV08}.
Assume that function
\verb+findMin+ has been verified for arbitrary integers. When
encountering a call to \verb+findMin+, CPBPV first checks if its
precondition is entailed by the constraint store, which requires a
consistency check of the constraint store with respect to the negation
of the precondition. Then CPBPV replaces the call by the
post-condition where the formal parameters are replaced by the actual
variables. In particular, for the first iteration of the loop and an
array length of 40, CPBPV generates the conjunction
$
0 \leq k^0 < 40 \; \wedge \; t^0[k^0] \leq t^0[0] \; \wedge \; \ldots \; \wedge \; t^0[k^0] \leq t^0[39]
$
which features {\tt element} constraint
\cite{VanHentenryck89}. Indeed, $k^0$ is a variable and a constraint
like $t^0[k^0] \leq t^0[0]$ indexes the array $t^0$ of variables using
$k^0$.

The modular verification of the selection sort explores only a single
path, is independent of the integer representation, and takes less
than $0.01s$ for arrays of size 40. The bottleneck in verifying
selection sort is the validation of function \verb+findMin+, which
requires the exploration of many paths. However the complete
validation of selection sort takes less than 4 seconds for an array of
length 6. Once again, this should be contrasted with the
model-checking approach of Eureka \cite{ABM07}. On a version of
selection sort where all variables are assigned specific values
(contrary to our verification which makes no assumptions on the
inputs), Eureka takes 104 seconds on a faster machine. Reference
\cite{ABM07} also reports that CBMC takes 432.6 seconds, that BLAST
cannot solve this problem, and that SATABS \cite{CKS05} only verifies
the program for an array with 2 elements.

\paragraph{\bf Sum of Squares}

Our last benchmark is described in \cite{CRV08}
and computes the sum of the square of the $n$ first integers stored in
an array. The precondition states that $n$ is the size of the array
and that $t$ must contain any possible permutation of the $n$ first
integers.  The postcondition states that the result is
$n\times(n+1)\times(2\times n+1)/6$.  The benchmark illustrates two
functionalities of constraint programming: the ability of specifying
combinatorial constraints and of solving nonlinear problems. The
\verb+alldifferent+ constraint\cite{Reg94} in the pre-condition
specifies that all the elements of the array are different, while the
program constraints and postcondition involves quadratic and cubic
constraints. The maximum instance that we were able to solve with
CPBPV was an array of size 10 in 66.179s.

CPLEX, the MIP solver, plays a key role in all these benchmarks. For
instance, the CP solver is never called in the Tritype benchmark. For
the Binary search benchmark, there are length calls to the CP solver
but almost 75\% of the CPU time is spent in the CP solver. Since there
is only path in the Buble sort benchmark, the CP solver is only called
once. In the Sum of squares example, 80\% of the CPU time is spent in
the CP solver.

\section{Discussion and Related Work} 
\label{related}

We briefly review recent work in constraint programming and model
checking for software testing, validation, and verification. We
outline the main differences between our CPBPV framework and existing
approaches.

\paragraph{\bf Constraint Logic Programming}

Constraint logic programming (CLP) was used for test generation of
programs (e.g., \cite{GBR98,JaV00,SyD01,GLM08}) and provides a nice
implementation tool extending symbolic execution techniques
\cite{BGM06}. Gotlieb et al. showed how to represent imperative
programs as constraint logic programs and used predicate abstraction
(from model checking) and conditional constraints within a CLP
framework. Flanagan \cite{Fla04} formalized the translation of
imperative programs into CLP, argued that it could be used for bounded
model checking, but did not provide an implementation.  The
test-generation methodology was generalized and applied to bounded
program verification in \cite{CoR06,CoR07}. The implementation used
dedicated predicate abstractions to reduce the exploration of spurious
execution paths. However, as shown in the paper, the CPBPV verifier is
significantly more efficient and often avoids the generation of
spurious execution paths completely.

\paragraph{\bf Model Checking}

It is also useful to contrast the CPBPV verifier with model-checking
of software systems. SAT-based bounded model checking for
software\cite{CBR01} consists in building a propositional formula 
whose models correspond to execution paths of bounded length violating
some properties and in using SAT solvers to check whether the
resulting formula is satisfiable. SAT-based model-checking platforms
\cite{CBR01} have been widely popular thanks to significant
progress in SAT solvers. A fundamental issue faced by model checkers
is the state space explosion of the resulting model. Various
techniques have been proposed to address this challenge, including
generalized symbolic execution (e.g., \cite{KPV03}), SMT-based model
checking, and abstraction/refinement techniques. SMT-based model
checking is the idea of representing and checking quantifier-free
formulas in a more general decidable theory
(e.g. \cite{GHN04,DuM06,NOR07}). These SMT solvers
integrate dedicated solvers  and share some of the motivations
of constraint programming. Predicate abstraction is another popular
technique to address the state space explosion. The idea consists in
abstracting the program to obtain an abstract program on which model
checking is performed.  The model checker may then generate an
abstract counterexample which must be checked to determine if it
corresponds to a concrete execution path.  If the counterexample is
spurious, the abstract program is refined and the process is
iterated. A successful predicate abstraction consists of abstracting
the concrete program into a Boolean program (e.g.,
\cite{BPR01,CKL04,CKS04}). In recent work \cite{AMP06,ABM07}, Armando
\& al proposed to abstract concrete programs into linear programs and
used an abstraction of sets of variables and array indices. They
showed that their tool compares favourably and, on some of the
programs considered in this paper, outperforms model checkers based on
predicate abstraction.\\
Our CPBPV verifier contrasts with SAT-based model checkers, SMT-based
model checkers and predicate abstraction based approaches: It does not
abstract the program and does not generate spurious execution
paths. Instead it uses a constraint-solver and nondeterministic
exploration to incrementally construct abstractions of execution
paths. The abstraction uses constraint stores to represent sets of
concrete stores. On many bounded verification benchmarks, our
preliminary experimental results show significant improvements over
the state-of-the-art results in \cite{ABM07}.  Model checking is well
adapted to check low-level C program and hardware applications with
numerous Boolean constraints and bitwise operations: It was
successfully used to compare an ANSI C program with a circuit given as
design in Verilog \cite{CKL04}. However, it is important to observe
that in model checking, one is typically interested in checking some
specific properties such as buffer overflows, pointer safety, or
user-specified assertions. These properties are typically much less
detailed than our post-conditions and abstracting the program may
speed up the process significantly. In our CPBPV verifier, it is
critical to explore all execution paths and the main issue is how to
effectively abstract memory stores by constraints and how to check
satisfiability incrementally. It is an intriguing issue to determine
whether an hybridization of the two approaches would be beneficial for
model checking, an issue briefly discussed in the next
section. Observe also that this research provides convincing evidence
of the benefits of Nieuwenhuis' challenge \cite{NOR07} aiming at
extending SMT\footnote{See also \cite{ABJ07} for a study of the
  relations between constraint programming and Satisfiability Modulo
  Theories (SMT)} with CP techniques.
 
\section{Perspectives and Future Work}
\label{conclusion}

This paper introduced the CPBPV framework for bounded program
verification.  Its novelty is to use constraints to represent sets of
memory stores and to explore execution paths over these constraint
stores nondeterministically and incrementally. The CPBPV verifier
exploits the fact that, when variables and arrays are bounded, the
constraint store can always be checked for feasibility. As a result,
it never explores spurious execution path contrary to earlier
approaches combining constraint programming and predicate abstraction
\cite{CoR06,CoR07} or integrating SMT solvers and the
abstraction/refinement approach from model checking \cite{ABM07}. We
demonstrated the CPBPV verifier on a number of standard benchmarks
from model checking and program checking as well as on nonlinear
programs and functions using complex array indexings, and showed how
to perform modular verification. The experimental results demonstrate
the potential of the approach: The CPBPV verifier provides significant
gain in performance and functionalities compared to other tools. 

Our current work aims at improving and generalizing the framework and
implementation. In particular, we would like to include tailored,
light-weight solvers for a variety of constraint classes, the
optimization of the array implementation, and the integration of Java
objects and references. There are also many research avenues opened by
this research, two of which are reviewed now.

Currently, the CPBPV verifier does not check for variable overflows:
the constraint store enforces that variables take values inside their
domains and execution paths violating these constraints are thus not
considered. It is possible to generalize the CPBPV verifier to check
overflows as the verification proceeds. The key idea is to check before
each assignment if the constraint store entails that the value
produced fits in the selected integer representation and generate an
error otherwise. (Similar assertions must in fact be checked for each
subexpression in the right hand-side in the language evaluation order.
Interval techniques on floats \cite{BGM06} may be used to obtain conservative
checking of such assertions. 

An intriguing direction is to use the CPBPV approach for properties
checking.  Given an assertion to be verified, one may perform a
backward execution from the assertion to the function entry point. The
negation of the assertion is now the pre-condition and the
pre-condition becomes the post-condition. This requires to specify
inverse renaming and executions of conditional and iterative
statements but these have already been studied in the context of test
generation. 

\paragraph{\bf Acknowledgements}
Many thanks to Jean-François Couchot for many helps on the use of the {\em{Why}}
framework. 

\end{document}